# Real Time Visualization of Dynamic Magnetic Fields with a Nanomagnetic FerroLens

*Emmanouil Markoulakis[a]\*, Iraklis Rigakis[a], John Chatzakis[a], Antonios Konstantaras[a], Emmanuel Antonidakis[a]*

[a]*Technological Educational Institute of Crete, ComputerTechnology, Informatics & Electronic Devices Laboratory, Romanou 3, Chania, 73133, Greece*

ARTICLE INFO

ABSTRACT

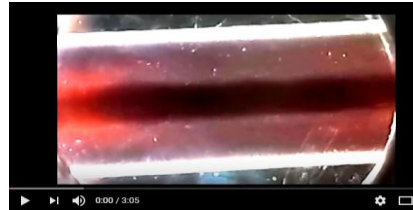

Fig.1 First Time Recorded in History EM Radio Wave inside Antenna ([Video Link])[4]

Due to advancements in nanomagnetism and latest nanomagnetic materials and devices , a new potential field has been opened up for research and applications which was not possible before. We herein propose a new research field and application for nanomagnetism for the visualization of dynamic magnetic fields in real-time. In short, Nano Magnetic Vision. A new methodology, technique and apparatus were invented and prototyped in order to demonstrate and test this new application. As an application example the visualization of the dynamic magnetic field on a transmitting antenna was chosen. Never seen before high-resolution, photos and real-time color video (Fig.1) revealing the actual dynamic magnetic field inside a transmitting radio antenna rod has been captured for the *first time*. The antenna rod is fed with *six hundred* volts, orthogonal pulses. This unipolar signal is in the very low frequency (*i.e. VLF*) range. The signal combined with an extremely short electrical length of the rod, ensures the generation of a relatively strong fluctuating magnetic field, analogue to the signal transmitted, along and inside the antenna. This field is induced into a ferrolens and becomes visible in real-time within the normal human eyes frequency observation. The name we have given to the new observation apparatus is, *SPIONs Superparamagnetic Ferrolens Microscope (SSFM)*, a powerful *passive* scientific observation tool with many other potential applications in the near future.

## 1. Introduction

Ever wondered what we would see if we could actually look with the naked eye at the dynamic magnetic field, part of the electromagnetic (EM) wave signal, on a transmitting radio antenna? In order to achieve this, we would have to invent an artificial magnetic sight apparatus and methodology that would enable us to observe the magnetic flux stream of the field under investigation. Similar to the natural *mechanism the migrating birds* [1] use, as researchers have indicated lately, that these birds may actually see the Earth's magnetic field in the sky propagating from geographic north, since their eyes act similarly to nano ferrohydrodynamic lenses. This magnetic sight apparatus, because our brain and eyes are not well suited for processing this kind of information, must be able to transform in real-time the magnetic flux field into an observable EM spectrum and correctly and adequately visualize all of its dynamic ( i.e. intensity, spatial and temporal) characteristics.

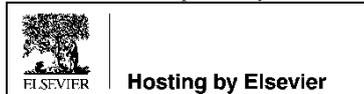

Hyperlinks:

1. http://www.ferrocell.us/intro.html
2. http://www.ferrocell.us/references.html
3. N42 Ring Magnet Data sheet:
   https://www.supermagnete.de/eng/data_sheet_R-27-16-05-N.pdf
4. New Discovery: First Time Recorded in History EM Radio Wave Video:
   https://www.youtube.com/watch?v=fGcvh4Rb0G4
5. Side by Side Comparison MIT Motion Amplification Algorithm video:
   https://www.youtube.com/watch?v=enTUapltGsA
6. Ferrofluid data sheet:
   https://ferrofluid.ferrotec.com/products/ferrofluid-educational-fluid/efh/efh1/

## 2. Materials and Methods

### 2.1. Basics

The ferrolens we use in our invention proposal is Timm Vandereli's patented and trademarked Ferrocell[TM1,2]. The basic operation principle of the ferrolens is illustrated in fig.2a,b,c,d. This nano Ferrohydrodynamic [2] Lens consists of a special mixture made utilizing, SPIONs [3] (i.e. superparamagnetic Iron Oxide Nanoparticles), to create a ferrofluid lens, a 62mm in diameter by 2mm thick, optical – magnetic device. Between the two 1mm thick round optical grade glass disks, is a 50μm thin active layer of superparamagnetic [4] ferrofluid. The glass and ferrofluid inside are sealed with an optical cement around the periphery to prevent the mixture from drying out, making up the final Ferrocell lens. See fig3.a (orange disk).

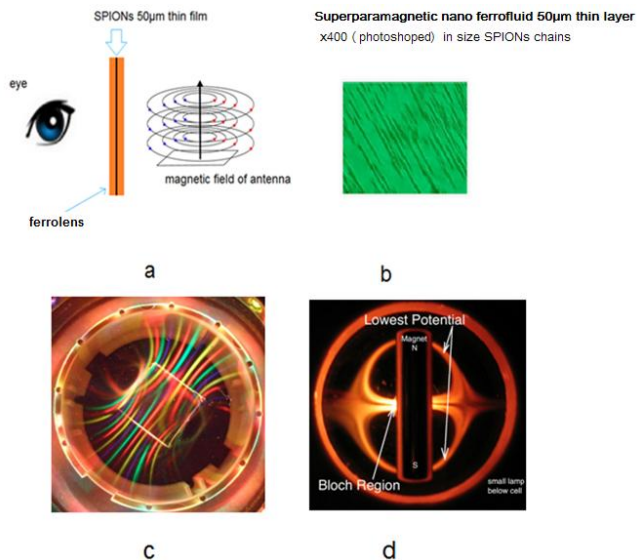

**Figure 2 (a)** eye looking through the Superparamagnetic 50μm thin SPIONs layer, placed inside the ferrolens, toward magnetic field of antenna **(b)** electronic microscope picture of the **10 nanometer** (nm) in size SPIONs forming dipole chains, times 400 enlarged. SPIONs absorb light in the visual spectrum thus are appearing colored and due to their superparamagnetic nature align with the magnetic flux lines of the antenna field effectively bending light and aligning it with the flux lines of the magnetic field therefore making it visible to the naked eye. **(c)(d)** Static magnetic field visualizations of permanent neodymium magnets as shown by the Ferrocell *(i.e. ferrolens).*

…………………………………………………………………………

In both cases, fig.2 (c)&(d), a light source is placed under the ferrolens, (c) a RGB LED strip peripherally  to the lens, (d) a small incandescent lamp. The ferrolens apparatus is then placed on top of the magnetic object its field we like to observe, in the case of fig.2(c)(d) the magnetostatic field of neodymium permanent magnets, a cube magnet (c) and (d) a cylindrical magnet. **Because the 10 nanometers (nm)** average in size SPIONs of the ferrofluid superparamagnetic properties *i.e. dipolar chains aligning to the magnetic field flux lines* [5] and also their *superabsorbing light properties* [6,7], the magnetic field becomes visible. Notice as well, the object itself under magnetic viewing due to light bending becomes transparent and practically invisible *i.e. invisibility cloak using nano ferrofluid* [8]. The color of the magnetic flux lines depends on the emitted light spectrum from the light source we use and the strength of the magnetic field. Generally, intense magnetic flux lines tend to attract more SPIONs creating thicker dipolar chains which  absorb colors in the red and yellow end of the spectrum, making them to appear blue-green and vice versa, smaller-sized SPION chains absorb blues and greens, resulting in a red appearance. Also, most importantly, absence of magnetic flux in a region of the magnetic field is depicted and displayed by the ferrolens as a dark region.

### 2.2 Dynamic Magnetic Field Operation

In the previous paragraphs we covered the basic operation of the superparamagnetic ferrolens under static conditions, whereas permanent magnets are used to induce a magnetic field, we then observed using the ferrolens and experimentally proved that the nano-sized SPIONs can effectively visualize  the intensity and spatial information of magnetic fields, so far so good. **But, what will  the results be when we try to observe a dynamic magnetic field,** such as, **the  fluctuating magnetic field on a transmitting radio antenna.** This has never been attempted before. Can the ferrolens adequately visualize and give us in real-time the temporal information of a dynamic, fluctuating, magnetic field?

For this lab experiment, a special prototype apparatus has been invented namely the SPIONs Superparamagnetic Ferrolens Microscope in short SSFM[©] described in fig3.(a)&(b)

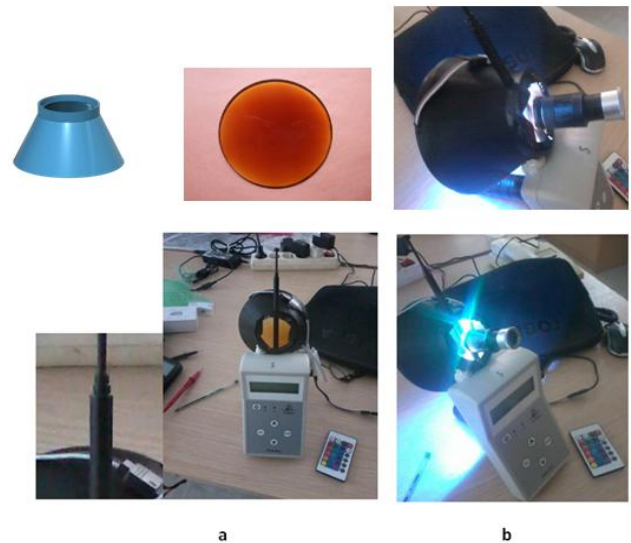

**Figure 3 (a) column** Programmable high voltage Pulse generator with the ferrolens assembly sitting on top and switched off. Orange-colored ferrolens placed above a 3D printer constructed lighting cone with a 24 RGB LED strip emitting *white light* attached to its inner bottom periphery controlled via infrared remote control. Black telescopic whip antenna rod under investigation, passing through the cone via holes and just 1mm under the ferrolens without any surface contact **(b) column** Ferrolens Microscope switched on and ready for use. An adjustable focus magnifying lens apparatus is placed on the top surface of the ferrolens with an N42 neodymium ring magnet in between, fastened under the magnifier and facing down the ferrolens center looking at the antenna rod behind the ferrolens.

To reach the  sensitivity threshold and improve responsiveness to magnetic field fluctuations  of the SSFM Microscope, a very electrically small [9], antenna rod is feed with a high voltage orthogonal unipolar 600 volts RMS signal using a programmable pulse generator fig.3(a)(b).

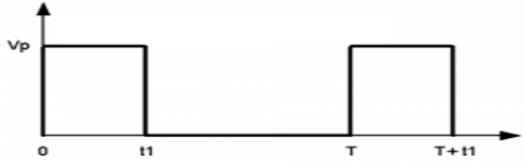

$$u_{RMS\_Pulse}{}^2 = \frac{1}{T}\int_0^{t1} Vp^2 \cdot dt = \frac{1}{T} \cdot Vp^2 \cdot t1|_0^{t1} = \frac{Vp^2}{T} \cdot t1$$

Thus,

$$u_{RMS\_Pulse} = Vp\sqrt{D} \tag{1}$$

Ideally from (1) above, we calculate for 600 volts RMS, $Vp = 900$ volts with a signal duty cycle equal to $D = 0.5$ with $D = t1/T$. Additionally, in order to overcome mass inertia on the SPIONs in the ferrofluid, thus, initially excite the ferrolens to improve sensitivity and further increase responsiveness of the superparamagnetic microscope, a N42 neodymium ring magnet is attached at the center of the ferrolens, top view, sandwiched between the ferrolens and the x2.4 magnifying lens sitting on top of the microscope, fig. 3(b) . Dimensions of the ring magnet[3] used, and fastened under the magnifying lens apparatus with exact axial alignment, are 26.75mm outer diameter x16mm inner diameter x5mm height, rated at Br = 1.3 Tesla of residual magnetism. *Therefore, the dynamic magnetic field strength generated by the signal applied at the transmitting antenna rod, is fluctuating at the microscope region on top of the static field induced by the ring magnet and is direct analogue to the signal and electromagnetic wave transmitted by the antenna.* Later on, we will also discuss the effect of the period T of the signal applied, thus, the effect of the signal's fundamental frequency $f_s$ on the *SSFM* microscope's sensitivity and responsiveness and prove the main criterion for choosing the operation frequency range of the proposed SSFM, nanotechnology SPIONs Superparamagnetic Ferrolens Microscope.

## 3. Results

### 3.1 A 'MRI' Scan of an Antenna

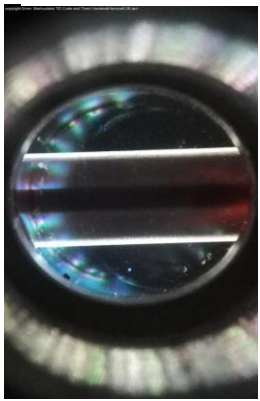

Microscopic Photography and video data information evidence *resembling a MRI scan*, were collected in real-time, of the actual magnetic field inside and around a transmitting antenna rod, *i.e. whip antenna*. A small sample of it is shown in fig.4 and fig.5(a)&(b) . All the recorded visual data evidence shown here, were taken looking through the 16mm inner diameter of the ring magnet attached above the ferrolens at its center with the aid of an additional magnification optical lens with adjustable focus and x2.4 zoom in. The telescopic whip antenna we use, has a maximum of 8mm thickness at the lower end, *is made out of bronze and is coated with black*

**Figure 4** Antenna rod is transparent under the powered up ferrolens *(A LED strip lighting is under the antenna rod)*

*ferromagnetic metallic paint. The later is very essential for the operation of the SSFM microscope since bronze is a diamagnetic* [10] *material and it would not be possible to look at the existing magnetic field inside the antenna rod unless it was coated with a ferromagnetic paint.* Thus, the antenna rod body becomes transparent (i.e looks like glass) while under magnetic viewing under the ferrolens shown at fig.4 and only the magnetic field inside and around of transmitting antenna is observed. In photographs fig.4 & fig.5 we can clearly observe (located at the center of the antenna body, in the middle), a black strip of accumulated unpolarized SPIONs nanoparticles population surrounding the Bloch domain wall region of the antenna's magnetic dipole field which is located across the width of the rod, and with the *Bloch region* [11] squeezed in the middle. A region where the magnetic field strength drops from nearly zero to zero across the few atoms thick Bloch region (shown also previously in fig2.d), a *quantum effect* shown by the ferrolens at *microscopic* level. Therefore *we specifically can say* that the observed black strip shown by the ferrolens of a transmitting antenna is effectively, the part of its magnetic field centered around the Bloch region. This is a region with zero magnetism (as discussed above, ferrolens colors zero magnetism as black). **This is a novel scientific observation and proof of its existence in antennas. To our knowledge this effect has never been reported or shown before, hereby demonstrating the potential capability of the SSFM microscope.**

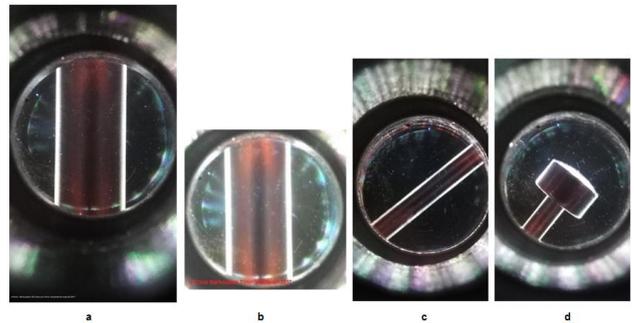

**Figure 5. (a)(b)** Transmitting antenna rod under the ferrolens. Field inside (red glow with black wall Bloch region) and outside the antenna rod (black and white magnetic field rays). A 'MRI' of an Antenna Rod. **(c)(d)** Fields at the thinner part of the whip (rod) antenna and the tip.

Continuing our analysis and observation of the field, in fig.5 (a)&(b) we observe two red-orange glowing halo bands, left and right of the centre part of the magnetic field (i.e. black strip in the middle) running along the total length of the antenna rod as we have observed. These reddish bands are more apparent when shown in fig.5 (c)&(d) as two reddish lines running in *parallel* along the antenna rod. These regions are in fact the **dipolar magnetic , North-South pole, fields** of the transmitting antenna. Notice also here the 'black sun' effect, distinct black finger rays separated by faint white rays, popping out from all directions around the antenna rod body, like the classical iron fillings experiment, specially noticeable around the periphery, these are the **radiation flux lines** of the antenna contained inside our ring magnet viewing window. **Specially, fig.5(d)** is an **amazing, first time, photograph showing off the antenna's tip magnetic dipolar field appearing like an upside down U-shaped horseshoe with a radiating magnetic black sun at the background.** The whole antenna structure resembles a magnetically vibrating tuning fork.

*3.2  A Dissection Analysis of the Field*

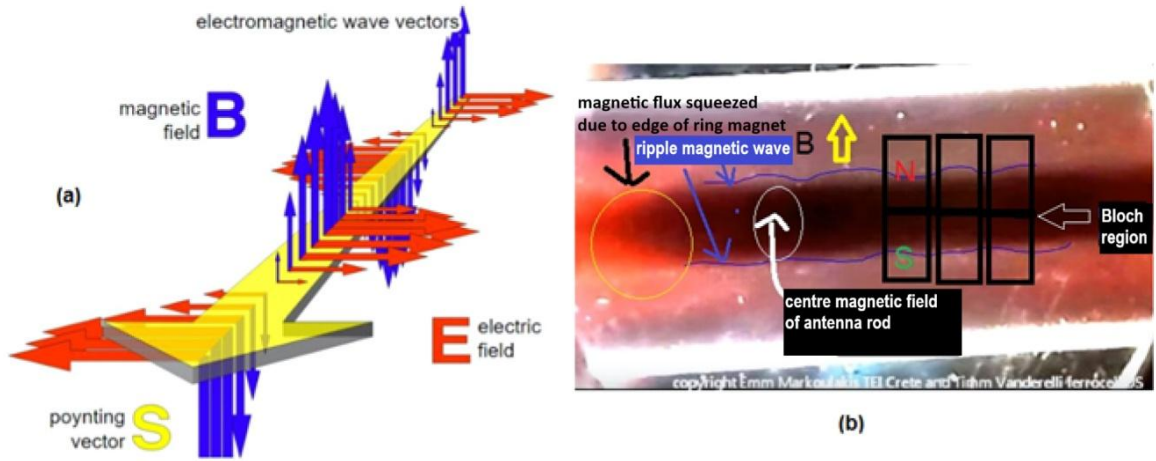

**Figure 6  Graphical analysis of the field**

In **Figure 6(a)** the classical graphical representation of Electromagnetic (EM) wave propagation [12] in space is shown, where **S** is the directional propagation Poynting vector of the wave and **B** and **E** the amplitude vectors of the wave's  magnetic and electric fields respectively. **Figure 6(b)** shows a superimposed graphics photograph version we have taken using the SSFM microscope of the magnetic field of a transmitting telescopic antenna rod (i.e. whip). By comparison of (a) and (b)  where the antenna rod length axis is pointing to the wave's  propagation S Poynting vector through space, we clearly can distinguish and identify in fig.6(b) the red halo bands above and below as the magnetic field, part of the Electromagnetic wave transmitted by the antenna, perpendicular to the EM wave propagation axis. To demonstrate this effect more **vividly** imagine the whole antenna rod length (*e.g. about 1 meter*) as a giant bar magnet with very large elongated North and South poles, similar to placing side by side in parallel ***multiple individual smaller magnets adding their magnetic flux lines coming from their North and South poles*** to form altogether the large in size magnetic dipole field of the transmitting, radiating antenna. Interestingly also, Fig6.(b) shows ***how all this Bloch wall regions of the many individual small magnets add up to form the total Bloch region of the magnetic field*** (i.e. black strip in the middle) of the antenna rod. Furthermore observation upon fig6(b), reveals that the red flux lines comprising the magnetic field appear more faint at the centre of our viewing window of  the microscope than they appear at the periphery where magnetic field looks more intense. This is due the interaction of the field of the antenna with the static field of the SSFM microscope's  ring magnet we used. The dynamic magnetic dipole field of the transmitting antenna is contained inside the static field of the ring magnet of the microscope, and gradually spatially squeezes the field of the antenna more and more until the antenna's flux becomes **V-*shaped*** close to the ring magnet where interaction is stronger fig6(b). Normally, without the microscope apparatus the magnetic dipole field of the antenna is ***extended outside its physical dimensions***.

*3.3  Dynamic Video Analysis of the Field and Motion Amplification*

Untill now in this paper we have presented a static analysis, by means of microscopic photography to record the observations  we obtained through the ferrolens, of the composite dynamic magnetic field of a transmitting antenna rod. However, in order to experimentally prove the SSFM microscope can efficiently display the dynamical temporal information data of the antenna field, a dynamic visual information time recording method was needed, thus, real-time high definition video recording is used. **We succeeded to record the first time ever video of the electromagnetic wave footprint on a transmitting antenna's dipolar magnetic field, Fig.1 Video Link[4] . In short, a magnetic wave fig 6(b).** The same result came out from all of our recordings. All of the original videos were recorder *at high-definition 1080p  without any video artifacts present*. Although the SSFM microscope cannot visualize and display electric field vectors  but only the magnetic part, it succeeded to register and display in *real-time* an EM wave signature of the signal transmitted by the antenna. Specifically, a periodic *sinusoidal transverse ripple effect* was observed upon the surface areas (*i.e. up and down surface*) of the antenna's magnetic field centre (*i.e. black strip in the middle of antenna*), projected clearly by its dynamic dipolar field (*i.e. red halo surrounding the centre field*) due to mainly  its envelope [13] amplitude fluctuations. Manifested in the ferrofluid, we speculate, as quantum effect phonons [14] or else called as ripplons by *Rosensweig* [15] *et al*. or Ferrofluid flows in AC and traveling wave magnetic fields by Markus Zahn [16] et al. and warrants further investigation. These ripples were constantly present along the whole length of the antenna rod with a frequency varying as measured from 1 to 5 Hertz approximately,  **Fig.1 Video Link[4]**, fig.6(b) and fig.7.

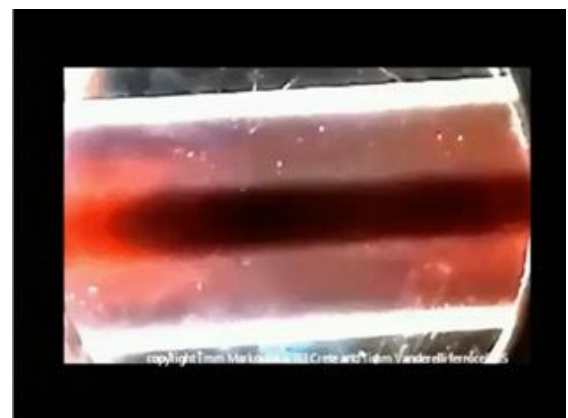

**Figure 7**  Animated gif of ripple effect (click)
https://drive.google.com/file/d/0B0A8uTBvEiQRTUlRYUFVckZRRHc/view

At first, this ripple motion of the field (because its microscopic nature), could not be detected by naked eye although there was a hint that something was in motion. Things changed dramatically after we processed the videos with **MIT Motion Amplification** [17] **Algorithm**, using the *Eulerian amplification* method. Motion amplification was set to **x20** at the frequency range *0.1 to 5 Hertz*. A high definition *side by side comparison video*[5] before and after the algorithm was applied, was also produced.

## 4. Discussion

At the end of section **2.2** of the paper, we have referred to the limitations in the operation of the ferrolens apparatus due to the maximum signal frequency *fs* at which the induced magnetic field, can be successfully displayed by the ferrolens, in phase (*i.e. in real-time*) with the applied signal. Thus, beyond and above this specific maximum frequency of operation, the ferrolens visual information display becomes progressively out of phase with the signal under observation, until for relative high field frequencies the SPIONs in the ferrofluid become totally unresponsive to the field fluctuations. Simply put, the SPIONs in the ferrofluid cannot keep up with the speed of the amplitude fluctuations of rapidly-changing dynamic magnetic field. *A criterion must be established for choosing the maximum safe operational frequency for the ferrolens $fs_{max}$.*
Rosensweig et al. [2,18–20] explains that the two dominant mechanisms through which the ferrofluid particle magnetic moment (*i.e. magnetic dipole movement* ) may align with the applied magnetic field are Brownian motion, which is the physical rotation of the particle into alignment with the field, and Néel relaxation, which is characterized by the movement of the particle magnetic moment relative to the crystal structure axis of the ferrofluid ferromagnetic particles.
Relaxation times for each are,

$$\tau_N = \frac{1}{f_o} e^{\frac{KV_N}{kT}} \quad \text{Néel relaxation time} \tag{2}$$

$$\tau_B = \frac{3V_B\eta_o}{kT} \quad \text{Brownian relaxaton time} \tag{3}$$

for which $\eta_o$ , $f_o$ , and $K$ are the carrier fluid dynamic viscosity, the frequency constant of Néel relaxation, and the anisotropy constant of the particle, respectively. The two particle volumes $V_B$ and $V_N$ are given by,

$$V_B = \frac{4}{3}\pi(R + \delta)^3 \quad \text{Brownian particle size} \tag{4}$$

$$V_N = \frac{4}{3}\pi R^3 \quad \text{Néel particle size} \tag{5}$$

In (4) $\delta$ represents the thickness of the adsorbed surfactant layer, and $R = d/2$ is the magnetic particle radius. The relaxation times defined in **(2) and (3)** are typically on the order of hundreds of milliseconds to nanoseconds [2,18]. The effective relaxation time for ferrofluid particles can be derived by considering that both the Brownian and Néel processes act simultaneously. When both mechanisms play a role in the relaxation process the **effective time constant** is,

$$\frac{1}{\tau_{eff}} = \frac{1}{\tau_B} + \frac{1}{\tau_N} \rightarrow \tau_{eff} = \frac{\tau_B\tau_N}{\tau_B+\tau_N} \tag{6}$$

A plot of the three *relaxation times as a function of particle diameter* [16,19] is shown in Figure 8, which indicates that the smallest time constant dominates the physical process of relaxation. For small particles Néel relaxation is faster than Brownian, and so the Néel time constant dominates $\tau_{eff}$. For large particles the Brownian relaxation is faster than Néel relaxation, and so the Brownian time constant dominates $\tau_{eff}$.

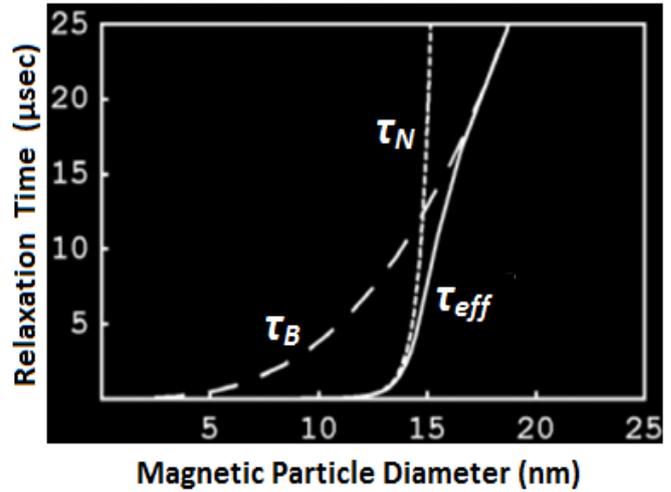

Figure 8 [2,20] The Brownian, Néel, and effective relaxation time constants as a function of spherical magnetic particle diameter. The plots correspond to EFH1 hydrocarbon-based ferrofluid, with the corresponding fluid parameters: the fluid mass density $\rho = 1169$ kg/m³, the dynamic viscosity $\eta_o = 10$ cP, the anisotropy constant $K = 23,000$ J/m for magnetite, the temperature $T = 300$ K, and the frequency constant, $f_o = 10^9$ Hz. The Brownian plot assumes zero surfactant thickness, $\delta = 0$.

Our ferrofluid (e.g. data sheet[6]) inside the ferrolens used in the experiments, is a type *EFH1 hydrocarbon-based* ferrofluid with an average of *magnetite SPIONs size, of 10 nanometers*, and with very similar specifications to the ones of the plot in fig.8. *Therefore, the plot in fig.8 is also true for our case and applies directly and we can securely say that our SSFM microscope can operate up to the tens or hundreds of Megahertz (MHz) frequency range.* Any *other movements* of the SPIONs inside the lens due to different phenomena other than the previously described are *ineffective or negligible* because the relatively strong existing **Van der Waals force** [21] within the ferrolens. The encapsulated **thin layer** of ferrofluid inside the ferrolens in this state, *does not flow anymore, but exists in a balanced state of equilibrium no matter what position the cell is oriented. The nanoparticles inside the ferrolens do not settle with gravity.* More in detail, the anionic surfactant coating [22] on the nanoparticles keeps the particles from touching each other (i.e. *clumping or agglomeration*) in the free state when there is no external magnetic field present. Notice here that the generated Van der Waals forces in the ferrofluid are not attractive but due to steric repulsion [2], results to stabilization.

**Nevertheless,** in our proposal we derive a maximum safe operational frequency *$fs_{max}$* for the SSFM microscope where magnetic susceptibility $x(\omega)$, of the SPIONs nanoparticles of the ferrofluid inside the ferrolens system, does not start to fall dangerously with frequency, in which case *magnetic moment of the SPIONs, is in phase with the induced external magnetic field by the signal* in order to have a *real-time* display of the field . The *later* dictates us that we are interested in finding a maximum frequency where the **real part** (*i.e. in phase*) of the susceptibility of the system is stable before it is starting to drop with increasing frequency.

According to Debye theory [23,24] [Debye 1929] for low-field regime the real and imaginary components of the complex susceptibility are thus given respectively by,

$$\chi'(\omega) = \frac{\chi_0}{(1+\omega^2\tau^2)} \tag{7}$$

$$\chi''(\omega) = \frac{\chi_0\omega\tau}{(1+\omega^2\tau^2)} \tag{8}$$

where $\tau$ *is effective relaxation time* [24] of the system given by (2),(3) and (6) and $x_0$ is the initial value [25] of susceptibility of our system given by, $x_0 = Msat/(3k_BT)$ the initial value [26] of susceptibility (9). *Msat* is the magnetic saturation value of the ferrofluid in A/m units which is given for our system, in SI units, at 44 mT[6] equivalently to 35 *KA/m* units, $k_B$ is the Boltzmann constant $1.38 \cdot 10^{-23}$ J/K and $T$ the temperature. For room temperature at $T = 300$ Kelvin the initial susceptibility of the ferrofluid of our system is given by the manufacturer  at $x_0 = 2.62$ in SI units[6] .

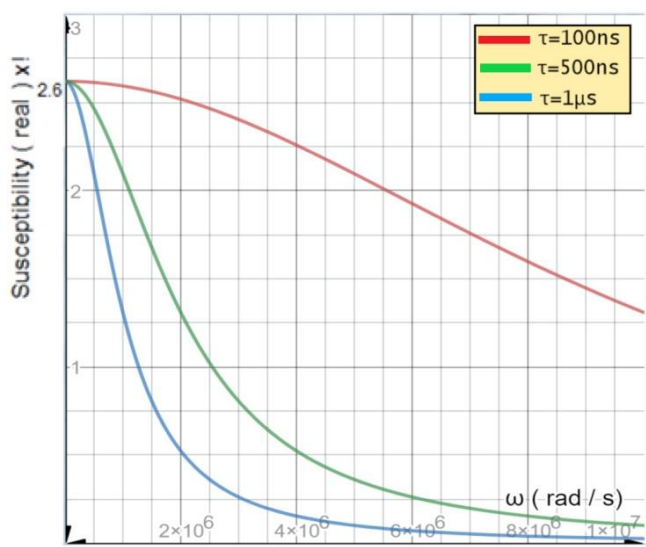

Figure 9  Susceptibility vs. frequency for various values of $\tau_{eff}$ system relaxation time. https://www.desmos.com/calculator/ha5n6qafbw

Using the information from (2),(3),(6),(7) equations the information from the data in fig.8 and the data from the ferrofluid manufacturer[6] we plotted in fig.9 the frequency dependency of our ferrolens's  magnetic susceptibility for three *worse case scenarios* values of $\tau_{eff}$  system relaxation times in order to draw our conclusions. From the above plot and considering the values for **real-time operation requirement** of the SSFM microscope *(i.e. there is no hysteresis in the display of the dynamic magnetic field fluctuations on the microscope)* and also considering as an **acceptable drop** for real-time operation, in the initial maximum value 2.62 (see fig.9) of the susceptibility, a *minus* 10%,  when calculated, $\textbf{\textit{fs}}_{max} = 40$ KHz maximum safe real-time operational frequency for 1μs (blue) relaxation time, 120 KHz for 500 ns relaxation (green) and  $\textbf{\textit{fs}}_{max} = 0.5$ MHz for 100 ns (red) relaxation time.

**However, we must stress here** that the above calculations are really worse case scenarios. In reality 10nm magnetite SPIONs EHF1 type ferrofluids encapsulated in the ferrolens glass structure (i.e. vacuum sealed μm thin layer of ferrofluid) are at the **10 ns relaxation time range** [27] and depending on the concentration percentage in the ferrofluid by the manufacturer, which gives us a safe real-time operating frequency as we have defined it, around 5 MHz as we  calculated https://www.desmos.com/calculator/pv1fzk6qm8  and observed experimentally. Furthermore, when real-time is not a problem and examining single frequency fields and not composite harmonical rich fields, with todays high initial [25] susceptibility valued $x_0$ manufactured ferrofluids,  we can happily operate the ferrolens at the tens and even hundreds of megahertz depending the size of the nanoparticles.

Regarding, increasing the field strength sensitivity of the ferrolens, few tens [28] of  mT static magnetic (*see section 2.2*) field should be applied depending  saturation  magnetization  value  given  by  ferrofluid manufacturer[6] (i.e. in our case it was 44mT) before a signal induced field is applied. In this way we ensure that all nanoparticles in the ferrofluid are activated and polarized with the external field. Also the signal amplitude applied should be at the hundreds of volts range with a low current. In our experiments we used a custom programmable high voltage pulse generator 600 Volts rms mostly at the VLF frequency range. *All photographs  and videos in this paper were taken from a 7 KHz pulsed, signal induced field.*  As a *suggestion for future experimentation* regarding our subject is the  use of a *100K frames per second  or more high speed camera*. *In this way someone  could video record* and then playback  in *slow motion* the full display effect  of the on-off  field transitions, induced by the pulsed signal.

## 5. Conclusion

**A new application for nanomagnetism is proposed.**  A new cost effective  nanotechnology  optical-magnetic  visualization,  passive technique, methodology and apparatus  *(i.e. SSFM microscope),* was invented, introduced and demonstrated using the superparamagnetic properties of a commercially available ferrolens. A micron thin film of ferrofluid encapsulated in a vacuum inside a lens  and applied to a custom-made optical microscope namely,  *"SPIONs Superparamagnetic Ferrolens Microscope"* or in short SSFM for  the observation and research of dynamic magnetic fields, which was described and prototyped in a lab while the operational parameters and limitations were analyzed and proven theoretically  and  by experiment. In order to demonstrate our results we applied the above apparatus and method, for visualizing the dynamic field inside and around a transmitting Radio Antenna rod and to our knowledge this has never been done before in this particular way. It successfully resolved and visualized all the spatial and temporal information of the dynamic field induced by the transmitter in the radio antenna signal with sensitivity factor down to nT scale and with real-time responsiveness up to 5 MHz observed by lab experiments i.e. limitations of our testing instrumentation prevented us from going higher in frequency. It was proven that in active thin layer only microns thick films of  ferrofluid when introduced inside the middle of an optical lens and illuminated by an artificial  light source, under the right conditions, have enhanced superparamagnetic and magnetic viewing properties compared to the free state (i.e. un-enclosed) ferrofluid, produces detailed visual information of the field in real-time and in color. Also it has been shown theoretically that with modern ferrofluid products, when real-time display is not an issue and for single frequency fields with little to no harmonics, the ferrolens can be used up to the *hundreds of MHz* range for SPIONs nanoparticles less than 10 nm in size. In addition the special optical properties of this kind of optical magnetic lenses were discussed and an invisibility cloaking as a side effect was presented (fig.4, fig2c). As a final remark, we believe that due the rapid scale-down of nanotechnology over the last years, nanomagnetism [29] will lead to more and more advanced ferrohydrodynamic products and methods with new potentials that before were not possible. This enables us to open up the view of our physical world from the macroscopic into the microscopic quantum effects scale. Additional research shall yield further development and novel applications of these advanced new nanomagnetic viewing devices.

## Acknowledgements

Foremost, we thank Timm Vanderelli of FERROCELL.USA, inventor of the patented Ferrocell™, for his resourceful technical help and support. Special thanks for Mr. Zisis Makris (research assistant) for his lab assistance in the experiments undertaken and also for assisting in the shooting of all the photographic and video recordings at the experiments. Thanks also to Dr. Stellios Kouridakis academic staff member and Mr. Vaggelis Tzavopoulos technical staff member for their scientific and engineering support.

## Appendix

Supplemental sample of experimental data information and visual material:

    I.    Inside google drive 560MB

https://drive.google.com/drive/folders/0B0A8uTBvEiQRT2p0dHRVa2VjWUE?usp=sharing

    II.    At https://ferrocellmicroscope.blogspot.gr/

.